\documentclass[aps,prd,preprintnumbers,superscriptaddress,nofootinbib,floatfix,twocolumn,notitlepage,amsmath,amssymb]{revtex4-2}

\usepackage{multirow}
\usepackage{amsmath}
\usepackage[utf8]{inputenc}
\usepackage{graphicx}
\usepackage{dcolumn}
\usepackage{graphicx}
\usepackage{xcolor}
\usepackage{slashed}
\usepackage{xspace}
\usepackage[utf8]{inputenc}
\usepackage{float}
\usepackage{subfigure}
\usepackage{url}
\usepackage{color}

\usepackage[dvipsnames,table]{xcolor}
\usepackage{comment}
\usepackage{adjustbox}
\usepackage{booktabs} 
\usepackage{colortbl} 

\def\beq#1\eeq{\begin{align}#1\end{align}}

\newcommand{\Bbar}{\,\overline{\!B}}

\RequirePackage{xspace}
\def\Bbar    {\kern 0.18em\overline{\kern -0.18em B}{}\xspace}

\definecolor{BlueViolet}{rgb}{0.2, 0.00, 0.7}
\definecolor{Blue}{rgb}{0.15, 0.00, 0.9}
\definecolor{halayaube}{rgb}{0.4, 0.22, 0.33}
\definecolor{sanddune}{rgb}{0.59, 0.44, 0.09}

\usepackage[colorlinks=true,linkcolor=Blue,citecolor=Blue,urlcolor=BlueViolet]{hyperref}
\usepackage[hyphenbreaks]{breakurl} 

\begin{document} 

\title{\bf Confronting $b \to (s, d)\mu^+ \mu^-$ and 
 $\mathcal{L}_{K^{*}\bar{K}^{*}}$ puzzle in a $U(1)_{B-3L_\mu}$ framework with neutrino phenomenology }

\author{Dhiren Panda}
\email{pandadhiren530@gmail.com}

\author{Priya Mishra}
\email{mishpriya99@gmail.com}

\author{Manas Kumar Mohapatra}
\email{manasmohapatra12@gmail.com}

\author{Rukmani Mohanta}
\email{rmsp@uohyd.ac.in}
\affiliation{School of Physics,  University of Hyderabad, Hyderabad-500046, India}


\begin{abstract}
\noindent
Current data on several observables in $b \to (s,d)$ $\mu^+ \mu^-$ transitions reveal certain discrepancies with Standard Model (SM) predictions. Considering the possibility that these deviations stem from new physics, we explore a $Z^{\prime}$ framework based on $U(1)_{B-3L_\mu}$ model. The parameter space is examined in light of existing constraints from $b \to (s,d)$ $\mu^+ \mu^-$ transitions. Within this framework, we show that the quark couplings accommodating $b \to (s,d)$ $\mu^+ \mu^-$ anomalies, imprint correlated signatures on the non-leptonic puzzles $\mathcal{L}_{K^{*}\bar{K}^{*}}$ in $B_{s(d)} \to K^{{*}}\bar{K}^{{*}}$ decays. We find that the scenario with both vector and axial-vector contributions of opposite sign  is favored by the data compared to pure vector contribution, offering a correlated probe of the underlying new-physics scenario. We further explore the neutrino phenomenology within the same framework in light of current data.
\end{abstract}
\maketitle
\section{Introduction}
Over the past three decades, precision studies of meson decays, particularly those involving $B$ mesons continue to play a crucial role in testing the Standard Model (SM) and probing potential signatures of new physics (NP). During this period, experimental efforts have achieved remarkable success in reducing the  uncertainties of the associated decay observables. At the same time, theoretical advances have been equally significant, with the development of various methods to account for QCD corrections, leading to increasingly refined SM predictions. In this regard, the flavor anomalies in $b \to s$ transitions provide one of the most intriguing potential indications of NP. This decay transitions have gained further support from the recent LHCb measurements sensitive to lepton flavor universality (LFU) violation. The key examples are the ratios $R_{K^{(*)}}$, $R_{K_S^{0}}$ and $R_{K^{*+}}$ \cite{Alguero:2022wkd, Alguero:2023jeh, Capdevila:2023yhq, Hurth:2023jwr},
\begin{equation}
\small R_X = \frac{\text{Br}(B \to X\, \mu^+ \mu^-)}{\text{Br}(B \to X\, e^+ e^-)}, 
\quad \text{where} \quad X \in \{K^{(*)},\, K_S^{0},\, K^{*+}\}.
\end{equation}
Additionally, the  well-known celebrated observable $P_5'$, has been precisely measured in recent years. The branching ratios of $B \to Y \mu^+\mu^-$ decays, with $Y=K^+,\, K^{0*},\, \phi$, as reported by the LHCb and CMS collaborations, exhibit notable deviations from the SM predictions. The details of such observables are listed in Table~\ref{Expt_Obs}. 

\begin{table*}[t]
\centering
\scriptsize
\renewcommand{\arraystretch}{1.05}
\setlength{\tabcolsep}{2.0pt}

\caption{Summary of $b \to s\ell^+\ell^-$ observables with SM predictions and experimental measurements in different $q^2$ bins.}
\label{tab:global_bsll}

\resizebox{\textwidth}{!}{
\begin{tabular}{lcccc}
\toprule

\textbf{Observable}
& \textbf{Quantity}
& \textbf{Bin 1}
& \textbf{Bin 2}
& \textbf{Bin 3} \\

\midrule

\multirow{3}{*}{$R_K$}

& $q^2$
& $[0.1,1.1]$
& $[1.1,6.0]$
& -- \\

& SM / Measurement
& $\begin{aligned}
1.00 \pm 0.01 \\
0.994^{+0.090+0.027}_{-0.082-0.029}
\end{aligned}$
& $\begin{aligned}
1.00 \pm 0.01 \\
0.949^{+0.042+0.023}_{-0.041-0.023}
\end{aligned}$
& -- \\

& Ref.
& \cite{Bordone:2016gaq,Hiller:2003js,LHCb:2022vje}
& same
& -- \\

\midrule

\multirow{3}{*}{$R_{K^*}$}

& $q^2$
& $[0.045,1.1]$
& $[1.1,6.0]$
& -- \\

& SM / Measurement
& $\begin{aligned}
1.00 \pm 0.01 \\
0.927^{+0.093+0.034}_{-0.087-0.033}
\end{aligned}$
& $\begin{aligned}
1.00 \pm 0.01 \\
1.027^{+0.072+0.027}_{-0.068-0.027}
\end{aligned}$
& -- \\

& Ref.
& \cite{Bordone:2016gaq,Hiller:2003js,LHCb:2022vje}
& same
& -- \\

\midrule

\multirow{3}{*}{$R_{K_S^0}$}

& $q^2$
& $[1.1,6.0]$
& --
& -- \\

& SM / Measurement
& $\begin{aligned}
1.00 \pm 0.01 \\
0.66^{+0.20}_{-0.14}{}^{+0.02}_{-0.04}
\end{aligned}$
& --
& -- \\

& Ref.
& \cite{Bordone:2016gaq,Hiller:2003js,LHCb:2021lvy}
& --
& -- \\

\midrule

\multirow{3}{*}{$R_{K^{*+}}$}

& $q^2$
& $[0.045,6.0]$
& --
& --
\\

& SM / Measurement
& $\begin{aligned}
1.00 \pm 0.01 \\
0.70^{+0.18}_{-0.13}{}^{+0.03}_{-0.04}
\end{aligned}$
& --
& --
\\

& Ref.
& \cite{Bordone:2016gaq,Hiller:2003js,LHCb:2021lvy}
& --
& --
\\

\midrule

\multirow{3}{*}{$R_\phi^{-1}$}

& $q^2$
& $[0.1,1.1]$
& $[1.1,6.0]$
& $[15,19]$ \\

& SM / Measurement
& $\begin{aligned}
1.016 \\
1.57^{+0.28}_{-0.25}\pm0.05
\end{aligned}$
& $\begin{aligned}
1.003 \\
0.91^{+0.20}_{-0.19}\pm0.05
\end{aligned}$
& $\begin{aligned}
1.002 \\
0.85^{+0.24}_{-0.23}\pm0.10
\end{aligned}$ \\

& Ref.
& \cite{Straub:2018kue,LHCb:2024rto}
& same
& same \\

\midrule

\multirow{4}{*}{$P_5^\prime$}

& $q^2$
& $[4.0,6.0]$
& $[4.3,6.0]$
& $[4.0,8.0]$ \\

& SM
& $-0.759\pm0.071$
& $-0.795\pm0.065$
& $-0.795\pm0.054$ \\

& Meas.
& $-0.439\pm0.111\pm0.036$
& $-0.96^{+0.22}_{-0.21}\pm0.25$
& $-0.267^{+0.275}_{-0.269}\pm0.049$ \\

& Ref.
& \cite{Descotes-Genon:2013vna,LHCb:2020lmf}
& \cite{Descotes-Genon:2012isb,CMS:2017rzx}
& \cite{Descotes-Genon:2014uoa,Belle:2016xuo} \\

\midrule

\multirow{3}{*}{$\mathcal{B}(B^0\to K^{*0}\mu^+\mu^-)$}

& $q^2$
& $[1.1,6.0]$
& --
& -- \\

& SM / Meas.
& $\begin{aligned}
(2.323\pm0.381)\times10^{-7} \\
(2.018\pm0.100\pm0.053)\times10^{-7}
\end{aligned}$
& --
& -- \\

& Ref.
& \cite{Bharucha:2015bzk,LHCb:2016ykl}
& --
& -- \\

\midrule

\multirow{3}{*}{$\mathcal{B}(B_s\to\phi\mu^+\mu^-)$}

& $q^2$
& $[1.1,6.0]$
& --
& -- \\

& SM / Meas.
& $\begin{aligned}
(2.647\pm0.319)\times10^{-7} \\
(1.41\pm0.073\pm0.024\pm0.068)\times10^{-7}
\end{aligned}$
& --
& -- \\

& Ref.
& \cite{Aebischer:2018iyb,Bharucha:2015bzk,LHCb:2021zwz}
& --
& -- \\

\midrule

\multirow{3}{*}{$\mathcal{B}(B_s\to\mu^+\mu^-)$}

& --
& --
& --
& -- \\

& SM / Meas.
& $\begin{aligned}
(3.672\pm0.154)\times10^{-9} \\
(3.361\pm0.028)\times10^{-9}
\end{aligned}$
& --
& -- \\

& Ref.
& \cite{Bobeth:2013uxa,Beneke:2019slt,Greljo:2022jac}
& --
& -- \\

\bottomrule
\end{tabular}
}
\label{Expt_Obs}
\end{table*}


Beyond semileptonic processes, rare non-leptonic $b \to s q\bar{q}$ ($q=u,d,s$) transitions provide an equally important and complementary probe of flavor-changing neutral currents (FCNC) effects through their sensitivity to short-distance dynamics encoded in hadronic observables. Owing to the interplay between weak interactions and nonperturbative QCD effects, these decays have been extensively investigated within the low-energy effective theory (LEFT)  and more recently, in the Standard Model Effective Field Theory (SMEFT). They have also served as sensitive probes of physics beyond the Standard Model, including scenarios with additional gauge bosons, scalar mediators, leptoquarks, supersymmetry, and other heavy degrees of freedom. The wealth of precision measurements of branching fractions, CP asymmetries, polarization observables, and isospin relations has established non-leptonic $b$-hadron decays as a powerful platform for testing the flavor sector and constraining the structure of new interactions.

The observed discrepancies in non-leptonic  $B$-meson decays have received relatively less attention compared to their semileptonic counterparts, despite their complementary sensitivity to flavor dynamics. One prominent example is the long-standing $B\to K \pi$ puzzle, which has been investigated for nearly two decades (see Refs.~\cite{Beaudry:2017gtw,Bhattacharya:2021shk} and references therein). In addition, tensions have recently been identified in several other hadronic decay observables, including the U-spin puzzle~\cite{Bhattacharya:2022akr}, three distinct discrepancies associated with the observables involving the $B_{d,s}^0\to K^{(*)0}\bar K^{(*)0}$ decay modes~\cite{Biswas:2023pyw}. Interestingly, this tension related to two-body decays of $B$ mesons into pairs of light pseudoscalar mesons, where $B={B^0,B^+,B_s^0}$ and $P={\pi,K}$. Motivated by these observations and the potential of hadronic observables to probe the nature of new physics beyond the Standard Model, in this work we focus our attention on the $B_{d,s}^0\to K^{*0}\bar K^{*0}$ decay channels.

While these anomalies have been interpreted within the quark sector alone, it is important to recognize that flavor structure of SM is not limited to only quarks, similar discrepancies are established in the lepton sector through neutrino oscillation experiments. Within the SM, neutrinos are massless, however, compelling evidence for nonzero neutrino masses and large mixing angles has emerged from solar~\cite{SNO:2002tuh}, atmospheric~\cite{Jung:2001dh}, reactor~\cite{KamLAND:2002uet}, and accelerator~\cite{K2K:2002icj} neutrino oscillation experiments. Extending the Standard Model with three right-handed neutrinos and allowing lepton number violation offers a natural explanation for the tiny, nonzero neutrino masses. In the present work, we extend the SM gauge symmetry by the local $U(1)_{B-3L_\mu}$ and include three right handed neutrinos for the requirement of gauge anomaly cancellation. We utilized type-I seesaw and Casses-Ibarra parametrization for the generation of tiny neutrino masses and mixing angles.

The paper is organized as follows. Section II presents the underlying theoretical framework, including the particle content of the model, the relevant interaction Lagrangians, and the construction Yukawa matrices in-line with the neutrino phenomenology. Section III is devoted to the signature of new physics in flavor sector associated with $b \to s(d) \ell^+ \ell^-$ quark level transition, with particular emphasis on the new physics in the non-leptonic $L_{K^*\bar{K}^*}$ puzzle. Finally, Section IV summarizes the main findings and presents our conclusions.

\section{Model Description}

We consider a simplified $U(1)$ extension of the Standard Model, wherein the SM  group has been  extended by an additional  $U(1)_{B-3L_{\mu}}$ local gauge symmetry to address  the anomalies in $B$ meson decays and neutrino phenomenology in a common platform.  

\begin{table*}[t]
\centering
\scriptsize

\setlength{\tabcolsep}{3pt}
\renewcommand{\arraystretch}{1.5}
\arrayrulewidth=0.1mm

\resizebox{\textwidth}{!}{
\begin{tabular}{||c||c|c|c|c|c|c|c|c|c||c|c||}
\hline \hline

\rowcolor{gray!30}
& \multicolumn{9}{c|}{\textbf{Fermions}}
& \multicolumn{2}{c||}{\textbf{Scalars}} \\
\hline

& $Q_L$
& $u_R$
& $d_R$
& $\ell_L$
& $e_R$
& $\psi_{d_{L,R}}$
& $N_{R_1}$
& $N_{R_2}$
& $N_{R_3}$
& $H$
& $\Phi$
\\

\hline

$SU(3)_C$
& 3
& 3
& 3
& 1
& 1
& 3
& 1
& 1
& 1
& 1
& 1
\\
\hline

$SU(2)_L$
& 2
& 1
& 1
& 2
& 1
& 1
& 1
& 1
& 1
& 2
& 1
\\
\hline

$U(1)_Y$
& $\frac{1}{6}$
& $\frac{2}{3}$
& $-\frac{1}{3}$
& $-\frac{1}{2}$
& $-1$
& $-\frac{1}{3}$
& 0
& 0
& 0
& $\frac{1}{2}$
& 0
\\
\hline

$U(1)_{B-3L_\mu}$
& $\frac{1}{3}$
& $\frac{1}{3}$
& $\frac{1}{3}$
& $(0,-3,0)$
& $(0,-3,0)$
& $\frac{1}{3}$
& 0
& -3
& 0
& 0
& 3
\\

\hline \hline

\end{tabular}
}

\caption{
Fields in the chosen $U(1)_{B-3L_\mu}$ model.
}

\label{model}

\end{table*}

 As the local $U(1)_{B-3L_\mu}$ gauge symmetry is not anomaly free with the  SM particle spectrum and requires additional fermions to be added. For generation dependent charge assignments, the $U(1))_{B-3L_\mu}$ symmetry requires the number of fermions and their charges depending on following anomaly cancellation, described as follows:
\begin{enumerate}
\item Pure gauge anomaly: $\rm \mathcal{A}^{SM}_{Q_{B-3L_\mu}^3} = 27  $
\item Mixed gauge \& gravitational anomaly:\\ $\rm \mathcal{A}^{SM}_{[\text{Grav}]^2 U(1)_{B-3L_\mu}} = 3$
\item Mixed gauge anomalies with SM gauge groups:
    \begin{align} \label{eq:2}
         \rm \mathcal{A}^{SM}_{[SU(3)_C]^2 U(1)_{B-3L_\mu}} &= 0, \quad \quad
        \rm \mathcal{A}^{SM}_{[SU(2)_L]^2 U(1)_{B-3L_\mu}} = 0.    \nonumber \\
         \rm \mathcal{A}^{SM}_{[U(1)_Y]^2 U(1)_{B-3L_\mu}} &= 0, \quad \quad
         \rm \mathcal{A}^{SM}_{[U(1)_{B-3L_\mu}]^2 U(1)_Y} = 0.
    \end{align} 
    \end{enumerate}
The non zero values of first two conditions enforced us to go beyond the SM particle content and include additional fermions.  We thus introduce three right-handed neutrinos (RHNs) ${N_{R_1}}, {N_{R_2}}, {N_{R_3}}$, for this purpose with general charges under the new gauge symmetry as ($Q_{N_1}, Q_{N_2}, Q_{N_3}$). The choice of these charges can be made in the following way from the first two anomaly conditions of Eq. (\ref{eq:2}):
\begin{align}
    Q_{N_1}^3+Q_{N_2}^3+Q_{N_3}^3 &= -27, \\
    Q_{N_1}+Q_{N_2}+Q_{N_3} &= -3.
\end{align}
The solution of these equations could be ($Q_{N_1}, Q_{N_2}, Q_{N_3}) = (a, -3,  -a)$, where $a \in (-R, ~R)$, could be a real number. The choice of `$a$' defines the possible interaction terms in the Lagrangian. In this work, we consider  the simplest choice as $a = 0$. 

For setting up an appropriate stage  for the correlative study, we additionally enrich the model with a vector-like down-type quark singlet $\psi_d$.  Alongside, a scalar singlet $\Phi$ is included to spontaneously break the new $U(1)$ gauge symmetry.   The complete list of field content along with their charges under various gauge groups are provided in Table.  \ref{model}.

\section{Neutrino Phenomenology}
In this section, we describe the model predictions for the neutrino sector. We start with the explanation of neutrino mass generation by employing type-I seesaw mechanism with the help of right-handed neutrinos. 

The relevant interaction Lagrangian for the lepton sector involving the right-handed neutrinos $N_{R_i}$  is given as

\begin{equation}
\begin{aligned}
\mathcal{L}\supset{}&
-y_{e1}\overline{L_e}\widetilde{H}N_{R_1}
-y_{e3}\overline{L_e}\widetilde{H}N_{R_3}
\\
&-y_{\mu2}\overline{L_\mu}\widetilde{H}N_{R_2}
-y_{\tau1}\overline{L_\tau}\widetilde{H}N_{R_1}
\\
&-y_{\tau3}\overline{L_\tau}\widetilde{H}N_{R_3}
\\[3pt]
&-\frac{M_{11}}{2}\overline{N_{R_1}^{c}}N_{R_1}
-\frac{M_{33}}{2}\overline{N_{R_3}^{c}}N_{R_3}
\\
&-\frac{M_{13}}{2}
\left(
\overline{N_{R_1}^{c}}N_{R_3}
+\overline{N_{R_3}^{c}}N_{R_1}
\right)
\\[3pt]
&-\frac{y_{12}}{2}\Phi
\left(
\overline{N_{R_1}^{c}}N_{R_2}
+\overline{N_{R_2}^{c}}N_{R_1}
\right)
\\
&-\frac{y_{23}}{2}\Phi
\left(
\overline{N_{R_2}^{c}}N_{R_3}
+\overline{N_{R_3}^{c}}N_{R_2}
\right)
+\mathrm{h.c.}
\end{aligned}
\end{equation}

with $\tilde{H}=i\sigma_2 H^*$, where $H$ is the SM Higgs doublet with vaccum expectation value as $\langle H \rangle=(0, v_H/\sqrt 2)^T $. The Dirac neutrino mass matrix is given as,
\begin{align}
M_D
=
\begin{pmatrix}
y_{e1} & 0 & y_{e3} \\
0 & y_{\mu2} & 0 \\
y_{\tau1} & 0 & y_{\tau3}
\end{pmatrix}
\frac{v_H}{\sqrt{2}}
\equiv
\mathbf{Y^\nu_{\alpha i}}\frac{v_H}{\sqrt{2}}.
\end{align}
Similarly, the Majorana interaction term leads to the RHN mass matrix,
\begin{align}
M_N
=
\begin{pmatrix}
M_{11} & m'_{12} & M_{13} \\
m'_{12} & 0 & m'_{23} \\
M_{13} & m'_{23} & M_{33}
\end{pmatrix}
\label{eq:MN}
\end{align}
where, $m'_{ij}$ = $y_{ij}\frac{v_\Phi}{\sqrt{2}}$, $v_\Phi$ being the VEV of the scalar field $\Phi$, i.e., $\langle \Phi \rangle = v_\Phi/\sqrt 2$. 

The Lagrangian for the charged leptons is written as,
\begin{equation}
\begin{aligned}
\mathcal{L}={}&
-y_{ee}\overline{e_L}He_R
-y_{e\tau}\overline{e_L}H\tau_R
-y_{\tau e}\overline{\tau_L}He_R
\\
&-y_{\mu\mu}\overline{\mu_L}H\mu_R
-y_{\tau\tau}\overline{\tau_L}H\tau_R
+\text{h.c.}
\end{aligned}
\end{equation}

In the charged-lepton sector, the Yukawa coefficients \(y_{\alpha\beta}\) are assumed to be real. This assumption does not reduce the generality of the framework since any unphysical complex phases can be eliminated through appropriate field redefinitions. After electroweak symmetry breaking, the Yukawa Lagrangian generates the charged-lepton mass matrix,
\begin{align}
M_\ell =
\begin{pmatrix}
y_{ee} & 0 & y_{e\tau} \\
0 & y_{\mu\mu} & 0 \\
y_{\tau e} & 0 & y_{\tau\tau}
\end{pmatrix}
\frac{v_H}{\sqrt{2}}
\equiv
\mathbf{Y^\ell_{\alpha\beta}}\frac{v_H}{\sqrt{2}}.
\label{eq:Charged-lepton}
\end{align}

The matrix \(M_\ell\) is, in general, non-diagonal and can be brought to diagonal form through a bi-unitary transformation,
\begin{equation}
V_L^{\ell\dagger} M_\ell V_R^\ell
=
\mathrm{diag}(m_e,m_\mu,m_\tau),
\end{equation}
where \(V_L^\ell\) and \(V_R^\ell\) denote unitary matrices acting on the left- and right-handed charged leptons, respectively. The eigenvalues \(m_e\), \(m_\mu\), and \(m_\tau\) correspond to the physical charged-lepton masses. To relate the model parameters with observables, it is useful to employ basis-independent quantities constructed from \(M_\ell M_\ell^\dagger\). The determinant of this Hermitian combination satisfies
\begin{equation}
\det(M_\ell M_\ell^\dagger)
=
m_e^2\,m_\mu^2\,m_\tau^2,
\end{equation}
while its trace yields
\begin{equation}
\mathrm{Tr}(M_\ell M_\ell^\dagger)
=
m_e^2+m_\mu^2+m_\tau^2.
\end{equation}
In addition, the second invariant can be expressed as
\begin{equation}
\begin{aligned}
\frac{1}{2}\Big[
&\left(\mathrm{Tr}(M_\ell M_\ell^\dagger)\right)^2
-\mathrm{Tr}\left((M_\ell M_\ell^\dagger)^2\right)
\Big]
\\
&=m_e^2m_\mu^2+m_\mu^2m_\tau^2
+m_\tau^2m_e^2.
\end{aligned}
\end{equation}

These relations are a direct consequence of the algebraic properties of a general \(3\times3\) matrix. The numerical forms of the unitary matrices \(V_L^\ell\) and \(V_R^\ell\), corresponding to the benchmark Yukawa couplings listed in Table~\ref{tab:Yukawa-values}, are provided in Appendix~\ref{App:A}.

\begin{table}[h]
\centering
\begin{tabular}{|c|c|c|c|c|}
\hline
$y_{ee}$ & $y_{e\tau}$ & $y_{\mu\mu}$ & $y_{\tau e}$ & $y_{\tau\tau}$ \\
\hline
~-0.805872~ & ~0.294368~ & ~0.0607364~ & ~-0.520862~ & ~0.189887~ \\
\hline
\end{tabular}
\caption{Charged lepton Yukawa couplings $Y_{\alpha\beta}$ scaled by $10^{-2}$.}
\label{tab:Yukawa-values}
\end{table}
Employing the type-I seesaw mechanism, the effective light-neutrino mass matrix is obtained as
\begin{align}
M_\nu
=
-
M_D M_N^{-1} M_D^T.
\label{eq:numass}
\end{align}

For convenience, we introduce,
\begin{equation}
\kappa = \frac{M_\nu}{v_H^2}.
\end{equation}

The matrix \(\kappa\) is diagonalized according to
\begin{equation}
V^{\nu T}\kappa V^\nu
=
D_\kappa
=
\mathrm{diag}(\kappa_1,\kappa_2,\kappa_3),
\end{equation}
whereas the RHN mass matrix satisfies
\begin{equation}
M_N
=
U_M^* D_N U_M^\dagger,
\qquad
D_N
=
\mathrm{diag}(M_{N_1},M_{N_2},M_{N_3}).
\end{equation}

Using the Casas--Ibarra parametrization~\cite{Casas:2001sr}, the neutrino Yukawa matrix can be reconstructed as
\begin{equation}
\mathbf{Y^\nu_{\alpha i}}
=
U_M^*
D_{\sqrt{N}}
\mathbf{R}
D_{\sqrt{\kappa}}
V^{\nu\dagger},
\label{eq:CI}
\end{equation}
where \(D_{\sqrt{x}} \equiv \sqrt{D_x}\), and \(\mathbf{R}\) represents a complex orthogonal matrix obeying
\begin{equation}
\mathbf{R}^T\mathbf{R}= I.
\end{equation}

The matrix \(\mathbf{R}\) contains the unknown high-energy degrees of freedom and determines the structure of the neutrino Yukawa couplings. By performing a scan over the parameter space of \(\mathbf{R}\), benchmark configurations consistent with neutrino oscillation data from NuFIT~6.1~\cite{Esteban:2024eli} are identified. In particular, five representative sample points satisfying all current oscillation constraints are selected. The corresponding neutrino Yukawa matrices $\mathbf{Y^\nu_{\alpha i}}$ and orthogonal matrices $\mathbf{R}$ are summarized in Tables~\ref{tab:benchmark_Y_R}, as shown in Appendix~\ref{App:A}. We consider a hierarchical RHN spectrum $(M_{N_1},M_{N_2},M_{N_3}) =
(10^{11},\,10^{12},\,10^{13})~\text{GeV}$.

Assuming a $U(1)_{B-3L_\mu}$ breaking scale of
$v_\Phi \approx 10^{14}\,\mathrm{GeV}$, the mass of the new gauge boson
$Z'$ is given by
\[
M_{Z'} = g' Q_\Phi v_\Phi.
\]
For guage coupling $g' \approx 10^{-11}$ and $Q_\Phi = 3$, the resulting
$Z'$ mass is approximately $10^{3}\,\mathrm{GeV}$.

\section{Signature of New Physics in Flavor Sector}
\subsection{$b \to q \mu^+ \mu^-$ ($q= s,d$)}
The effective Hamiltonian governing the  quark level transition $b \to s \ell^+ \ell^-$, including contributions from possible new physics operators, can be expressed as~\cite{Buras:1994dj},

\begin{align}
{\cal H}_{\rm eff}
&=
-\frac{\alpha G_F}{\sqrt 2 \pi} V_{tb}V_{ts}^*
\Big[
2 \frac{C_7^{\rm eff}}{q^2}
\left(
\bar s \sigma^{\mu \nu} q_\nu (m_s P_L +m_b P_R) b
\right)
(\bar \ell \gamma_\mu \ell)
\nonumber\\
&
+ C_9^{\rm eff} (\bar s_L \gamma^\mu b_L)
(\bar \ell \gamma_\mu \ell)
+ C_{10} (\bar s_L \gamma^\mu b_L)
(\bar \ell \gamma_\mu \gamma_5 \ell)
\nonumber\\
&
+ {C}_{9}^{\ell \ell, NP}
(\bar{s}_L \gamma^{\mu} b_L)
(\bar{\ell}\gamma_{\mu}\ell)
+ {C}_{10}^{\ell \ell, NP}
(\bar s_L \gamma^{\mu} b_L)
(\bar{\ell}\gamma_{\mu}\gamma_{5}\ell)
\Big],
\label{Ham-SM}
\end{align}
where $C_7 ^ {\rm eff}, C_9 ^ { \rm eff}$, and $C_{10}$ are the Wilson coefficients  assessed at $\mu =m_b^ {\rm pole}$ scale \cite{Buras:1994dj}. $G_F$ is the Fermi coupling constant, $\alpha$ is the fine structure constant, and $V_{ij}$ are the CKM matrix elements. The effective coupling constants for the corresponding vector and axial vector NP operators are $C_9^{\ell \ell, NP}$ and $C_{10}^{\ell \ell, NP}$, respectively.

The Yukawa interactions between the vector-like fermions and their SM counterparts result in the $Z'$ effective couplings to the SM fermions in the study of $b \to q \mu^+ \mu^-$ transition decays. The parametrized couplings in the interaction terms, are provided as follows:
\begin{equation}
\mathcal L \supset \bar f_i \gamma^\mu \left( \lambda_L^{f_i f_j} P_L + \lambda_R^{f_i f_j} P_R \right) f_j Z_\mu^\prime \, .
\end{equation}
\begin{figure*}[t]
    \centering

    \begin{minipage}[t]{0.45\textwidth}
        \centering
        \includegraphics[width=\textwidth]{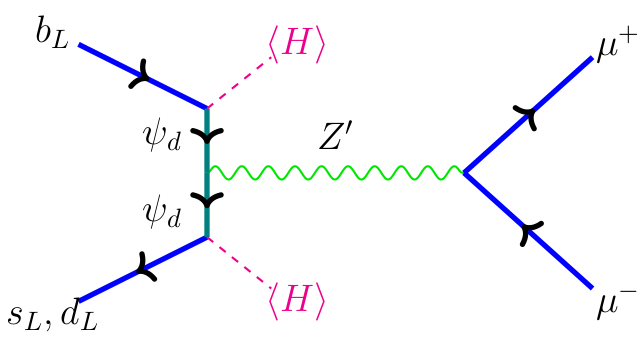}
    \end{minipage}
    \hfill
    \begin{minipage}[t]{0.45\textwidth}
        \centering
        \includegraphics[width=\textwidth]{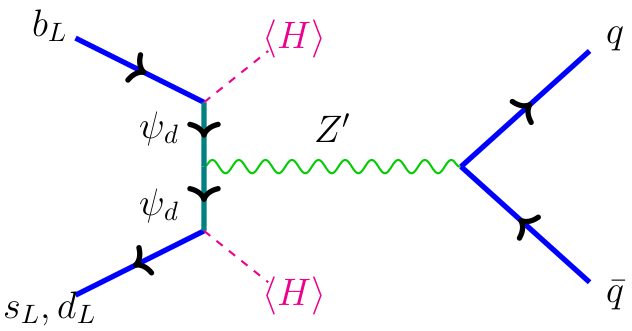}
    \end{minipage}

\caption{Generation of new physics operators via SM -- vector-like fermion mixing for (i) $b \to s \mu^+ \mu^-$ and (ii) $b \to s q \bar{q}$ transitions.}
\label{fig::RSigma}
\end{figure*}

In the $b \to s \ell^+ \ell^-$ transitions, the new physics is assumed to occur only in the muon sector. To solve the descripancies in $b \to s$ and $b \to d$ type exclusive semileptonic transition decays, the induced $Z^\prime - b - s$ and $Z^\prime - b - d$ couplings alongside $Z^\prime - \mu - \mu$ at tree level are found to be 
\begin{equation}
\lambda_L^{bs(d)} = \frac{1}{3} g_X [V_d]_{42(1)} [V_d]_{43} \,.  
\end{equation}
Meanwhile, the flavor-conserving couplings, $\lambda_L^{\mu\mu}$ and 
$\lambda_L^{q\bar{q}}$, remain unaltered. The associated Feynman diagrams are shown in Fig.\ref{fig::RSigma}.

Given our model, the Yukawa interaction with the vectror-like fermion $\psi_q$ and the scalar $\Phi$ couple only to the SM fermion $q_L$ ($\mu_L$). Since the SM fermions involved are purely left-handed, the operators $O_9^{\mu,NP}$ and $O_{10}^{\mu,NP}$ are in fact produced simultaneously in the presence of $\lambda_R^{\mu \mu}=0$, with their Wilson coefficients $C_9^{\mu \mu,NP}$ and $C_{10}^{\mu \mu,NP}$. These new physics couplings are expressed below \cite{Altmannshofer:2014rta}, 
\begin{equation} \label{NP coupling}
C_9^{\mu \mu, \, \text{NP}} = - C_{10}^{\mu \mu, \, \text{NP}} = -  \frac{\pi}{\sqrt{2} G_F \alpha m_{Z'}^2}  \frac{\lambda_L^{bs(d)} \, \lambda_L^{\mu \mu}}{V_{tb} V_{ts}^\ast} \,,
\end{equation}
where $\alpha$ and $G_F$ are fine structure constant and the Fermi coupling constant, respectively. Among the possible new physics scenarios in $Z'$ model, the most feasible environment to study the $b \to s \mu^+ \mu^-$ transitions is $C_9^{\mu \mu,NP}=-C_{10}^{\mu \mu,NP}$. Based on the available theoretical and experimental inputs, many works have been performed in this scanario \cite{Alok:2017sui, Greljo:2015mma, Chiang:2016qov, Crivellin:2015era, Mohapatra:2021ynn}. While models with the $Z'$ couplings theoretically motivate the scenario $C_9^{\mu \mu, NP} = -C_{10}^{\mu \mu, NP} < 0$, we also expand our analysis to include only the dominating contribution, $C_9^{\mu \mu, NP} < 0$. Furthermore, the global fits to $b \to s \mu^+ \mu^-$ transitions consistently favor a negative shift in $C_9^{\mu \mu, NP}$, which also motivates for the study of $b \to s \mu^+ \mu^-$ transitions \cite{Descotes-Genon:2015uva, Alguero:2019ptt, Capdevila:2017bsm, Alguero:2021anc}. In light of these two scanarios, we focus our analysis accordingly. Taking the ingredients at hand, as given in Eq. (\ref{NP coupling}), we will focus on the NP parameters $\lambda_L^{bs(d)}$ and $\lambda_L^{\mu \mu}$ with the $Z'$ boson mass 1 TeV, as discussed in detail in sub-section \ref{Ntp}. 


\subsection{Neutrino trident production}
\begin{figure}[h!]
\centering
\includegraphics[height=50mm,width=55mm]{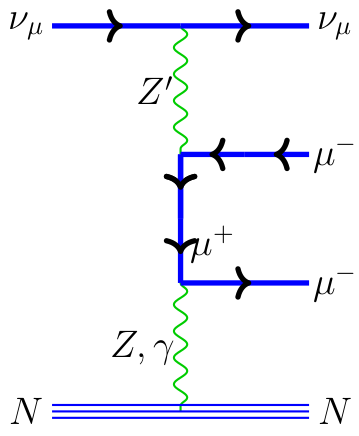}
\caption{Tree-level diagram illustrating $Z^{\prime}$
contribution to the neutrino trident production.}
\label{Neutrinotrident}
\end{figure}

The process $\nu_\mu N \to \nu_\mu N \mu^+ \mu^-$ involving muon pair generation in neutrino-nucleus scattering, serves as a sensitive probe for exploring NP scenarios \cite{Altmannshofer:2014pba}.
A $Z^{\prime}$ boson that interacts with muon neutrinos ($\nu_\mu$) can contribute to the process $\nu_\mu N \to \nu_\mu N \mu^+ \mu^-$, where $N$ denotes a heavy nucleus. One such contribution is illustrated in Fig. \ref{Neutrinotrident}. 
In the limit of a heavy $Z^{\prime}$, the cross section $\sigma$ for neutrino trident production, $\nu_{\mu} N \rightarrow \nu_{\mu} N \mu^+\mu^-$, can be represented as follows \cite{Alok:2017jgr}:
\vspace{-0.5em}
\begin{align} \label{trident}
R_\nu =  \frac{\sigma}{\sigma_{\rm SM}} 
=& \frac{1}{1+(1+4s^2_W)^2} \Bigg[\left(1+ \frac{v^2(g^{\mu\mu}_L)^2}{M^2_{Z'}}\right)^2 \nonumber\\
+& \left(1+4s^2_W+\frac{v^2(g^{\mu\mu}_L)^2}{M^2_{Z'}}\right)^2\Bigg],
\end{align}
where $v = 246$ GeV and $s_W = \sin\theta_W$. Here $g_L^{\mu \mu} = \frac{2 g_X |\lambda_{\psi _\ell}^ \mu|^2 v_\phi^2}{2 m_{\psi _\ell}^2 + |\lambda_{\psi _\ell}^ \mu|^2 v_\phi^2}$. In the present model, the $Z^{\prime}$ couples only to the left-handed muon current with coupling $g^{\mu\mu}_L$, with $g^{\mu\mu}_R=0$. We use the ratio $R_\nu = \sigma / \sigma_{\rm SM}$, which is defined in eq.~(\ref{trident}), to constrain the parameter $g_L^{\mu \mu}$. The measurement $R_\nu^{\rm Exp} = 0.82 \pm 0.28$, reported by the CCFR collaboration~\cite{CCFR:1991lpl}, serves as a valuable probe for testing new physics, offering a direct comparison against the SM prediction. With $g^{\mu \mu}_R=0$, the comparison yields an upper limit on $(g^{\mu \mu}_L)^2/m_{Z^{\prime}}^2$. Based on Drell--Yan production at ATLAS experiment~\cite{ATLAS:2024uvu}, $pp \to W^{\pm(*)} \to Z' \mu^\pm \nu \to \mu^\pm \mu^\mp \mu^\pm \nu$, a search for a $Z'$ gauge boson predicted by the $L_\mu - L_\tau$ model has been conducted. Similarly, the mass of the $Z'$ boson in the $L_\mu - L_\tau$ model is constrained by the ATLAS experiment using the process $pp \to Z' \mu \mu \to 4\mu$ \cite{ATLAS:2023vxg}. Recently, the authors Trocsanyi et al. \cite{Peli:2024fha} performed the exclusion limit based on the $B-L$ extension of the SM while taking into account the ATLAS and CMS direct searches \cite{ATLAS:2019erb, CMS:2021ctt}. Since the bounds on the $Z'$ mass and couplings are not well established for the $U(1)_{B - 3L_\mu}$ model, we consider three representative values of $m_{Z'}$ to explore the phenomenology and enable comparison with different scenarios. For a fixed $Z'$ mass of $1$ TeV, the corresponding coupling is $g_L^{\mu \mu}=1.137$. This benchmark value is adopted in our subsequent analysis.



\subsection{$B_q - \bar{B}_q$ mixing ($q=s,d$)}
\begin{figure}[h!]
\centering
\includegraphics[height=44mm,width=76mm]{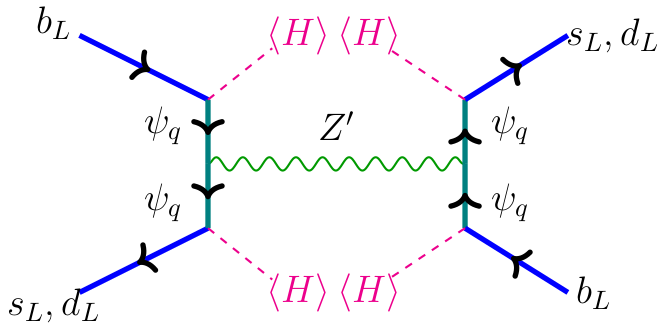}
\caption{The $Z^{\prime}$ contribution at tree-level illustrating $B_q - \bar{B}_q$ mixing.}
\label{Mixing}
\end{figure}

The SM contribution to the $B_q - \bar{B}_q$ mixing occurs at loop-level box diagram via $W$ boson and up-type quarks running in the loop with the dominant contribution from virtual top quark. However, as illustrated in Fig. \ref{Mixing}, the present model inducing the FCNC $b-q-Z'$ coupling influences the $B_q - \bar{B}_q$ mixing at the tree level. 
The equation for $M^q_{12}$, in the box diagrams responsible for mixing, is given as
\begin{equation} 
M^q_{12}= \frac{1}{3}M_{B_q}f_{B_q}^2 \widehat{B}_{B_q}\left[N\, C_{\rm VLL}^{\rm SM} + \frac{\left(g^{bq}_L\right)^2}{2M_{Z'}^2}\right], 
\end{equation} 
with 
\begin{equation}
\begin{aligned}
N &= \frac{G_F^2 M_W^2}{16\pi^2}
(V_{tb}V_{tq}^*)^2, \\[6pt]
C_{\rm VLL}^{\rm SM}
&= \eta_B x_t \Bigg[
1+\frac{9}{1-x_t}
-\frac{6}{(1-x_t)^2}
-\frac{6x_t^2\ln x_t}{(1-x_t)^3}
\Bigg],
\end{aligned}
\end{equation}
where $x_t \equiv m_t^2 / M_W^2$. Here, $\eta_B = 0.84$ is the short-distance QCD correction calculated at NNLO~\cite{Buchalla:1995vs}, $f_{B_q}$ is the $B$ meson decay constant, and $\widehat{B}_{B_q}$ is the bag factor.

To obtain the mass difference $\Delta M_q = 2|M_{12}^q|$, the following expression is used,
\begin{equation}
\Delta M_q = \Delta M_q^{\rm SM} \left|1 + \frac{(g^{bq}_L)^2}{2 N\, C_{\rm VLL}^{\rm SM} M_{Z'}^2} \right|\,.
\label{eq: Bdmix}
\end{equation}

The measured mass differences in the neutral $B_s$ and $B_d$ meson systems are given by
\begin{equation}
\begin{aligned}
\Delta M_s &= 17.757 \pm 0.020~{\rm ps}^{-1},\\
\Delta M_d &= 0.5065 \pm 0.00192~{\rm ps}^{-1},
\end{aligned}
\label{eq:mass_differences}
\end{equation}
where $\Delta M_s$ and $\Delta M_d$ denote the mass differences between the heavy and light mass eigenstates in the corresponding neutral $B$-meson systems~\cite{HFLAV:2016hnz}.
\begin{figure*}[t]
\centering

\includegraphics[width=8cm, height=8cm]{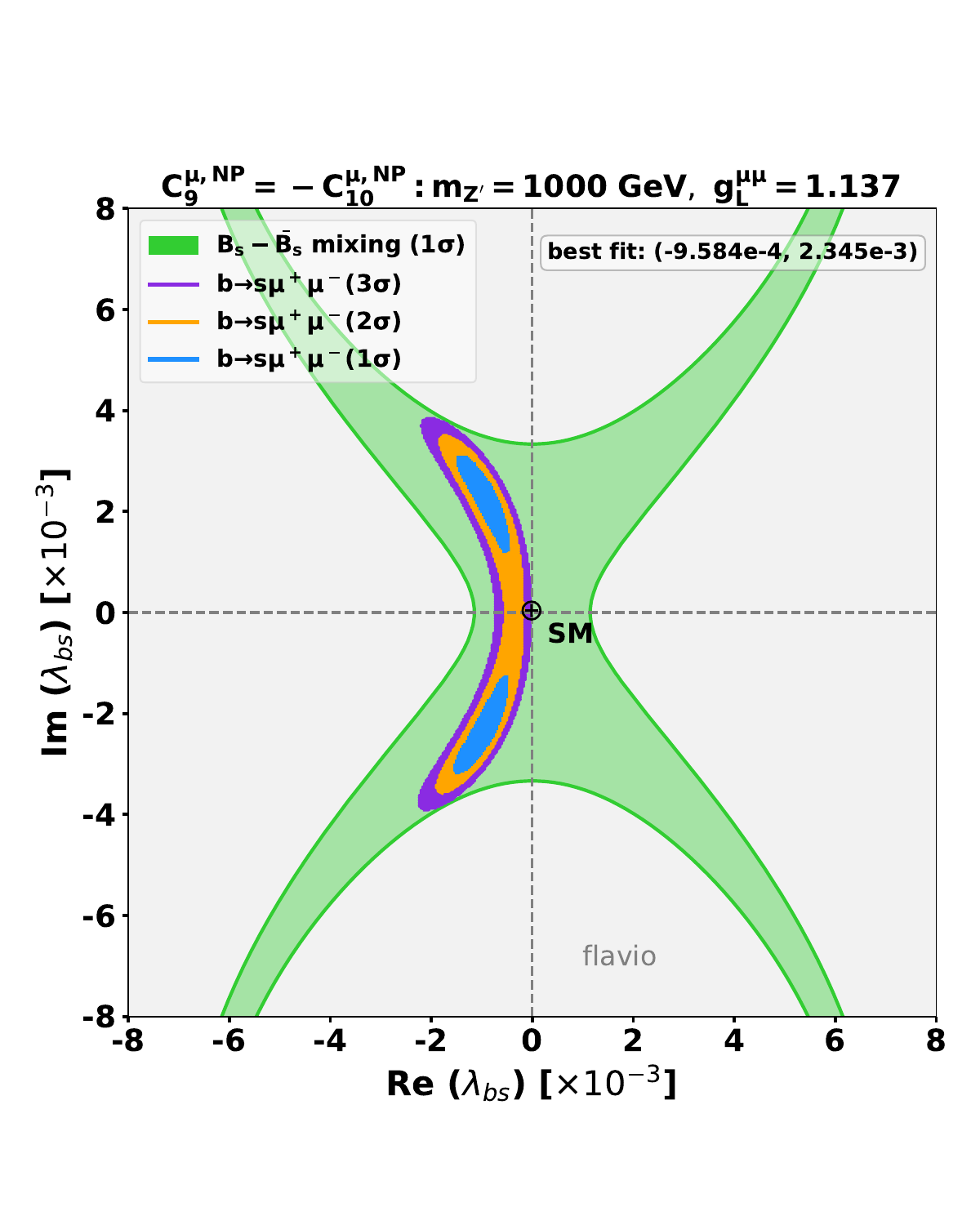}
\includegraphics[width=8cm, height=8cm]{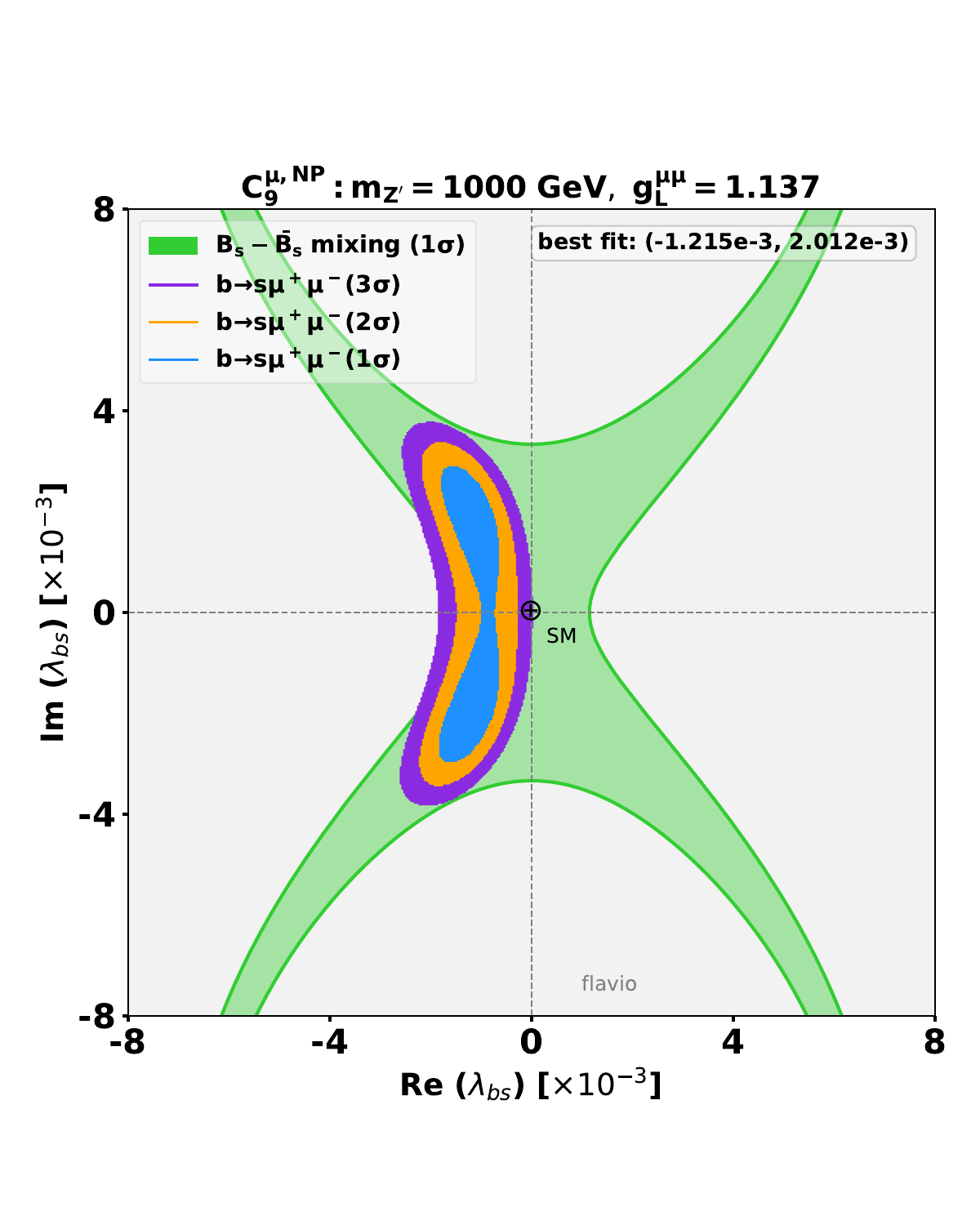}

\caption{
Parameter space scan in the $\lambda_{bs}^{\rm Re}$--$\lambda_{bs}^{\rm Im}$ plane:
$C_9^{\rm NP}=-C_{10}^{\rm NP}$ (left panel) and $C_9^{\rm NP}$ (right panel).
}

\label{2D:contour1}

\end{figure*}


\begin{figure*}[t]
\centering
    \includegraphics[width=8cm,height=8cm]{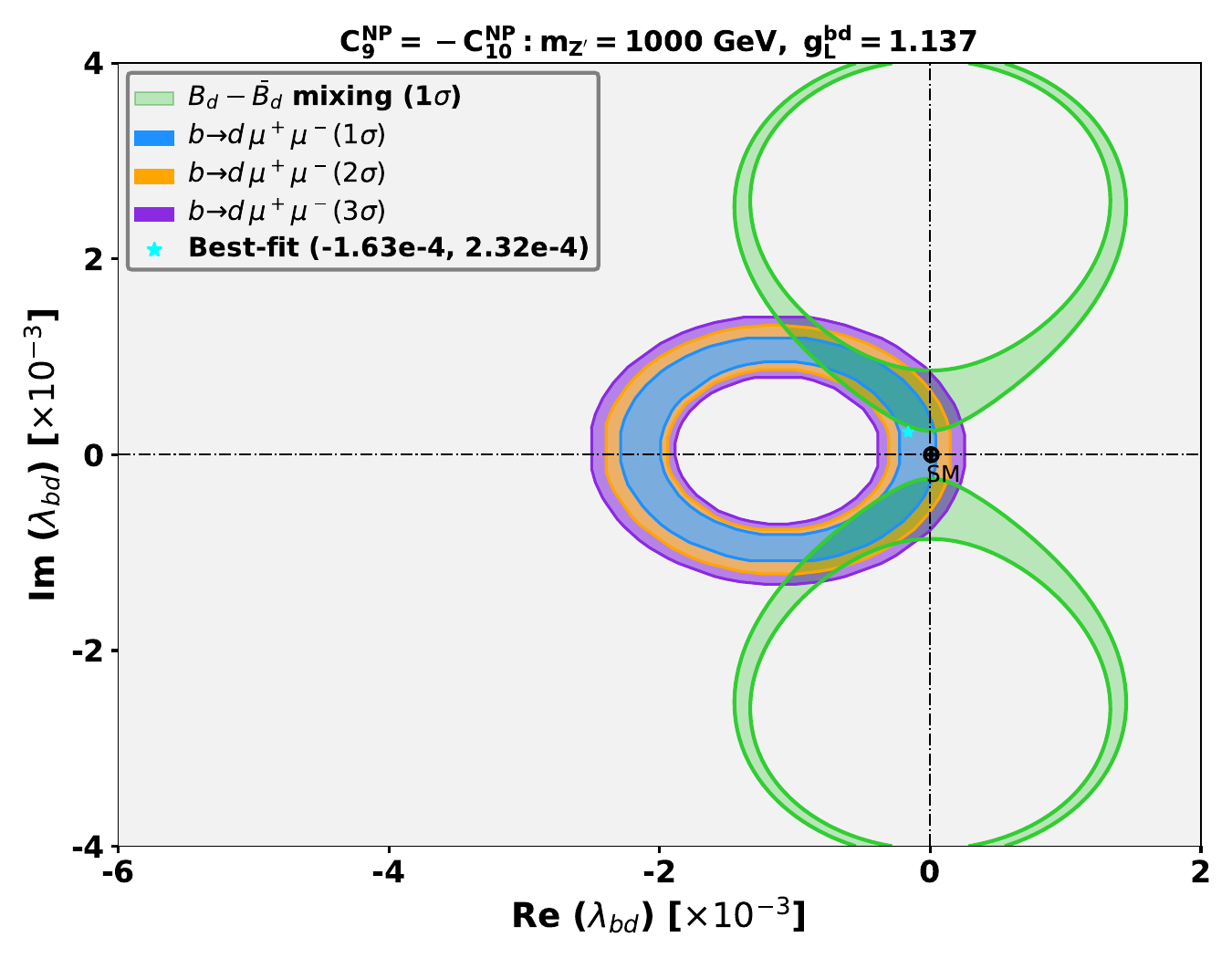}    
   \includegraphics[width=8cm,height=8cm]{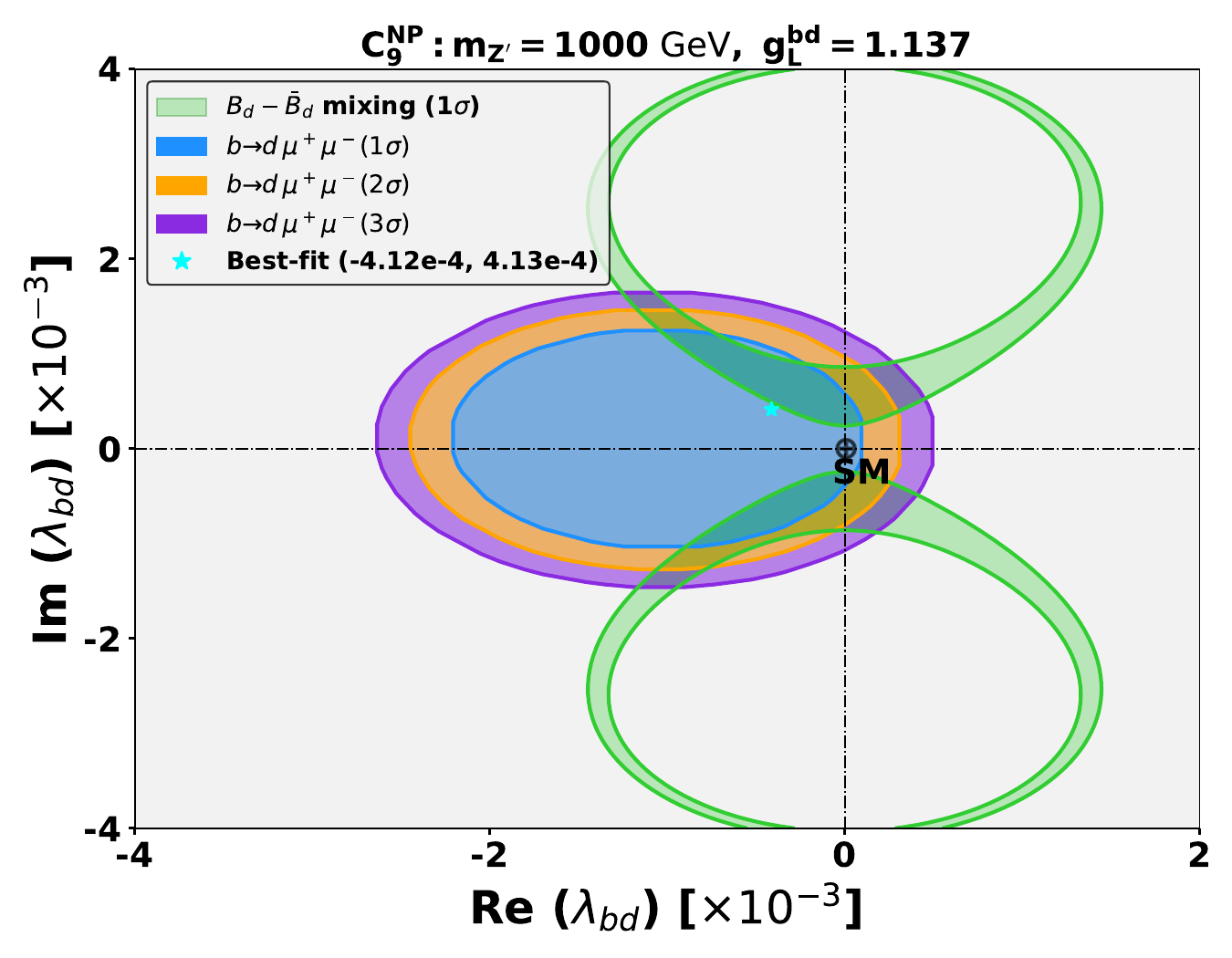}
\caption{Parameter space scan in $\lambda_{bd}^{\rm Re}$ - $\lambda_{bd}^{\rm Im}$ plane: $C_9^{\rm NP} = -C_{10}^{\rm NP}$ (left) and $C_9^{\rm NP}$ (right).}
\label{2D:contour2}
\end{figure*}

\subsection{Role of new physics parameter $\lambda_{bq}$ in semileptonic $b \to q \mu^+ \mu^-$ transitions} {\label{Ntp}}
To constraint the new physics parameter space, we perform the analysis in the complex $\lambda_{bs}$ plane, incorporating the constraints from $B_s-\bar B_s$ mixing and $b\to s\mu^+\mu^-$ observables. The resulting allowed regions are displayed in Fig. \ref{2D:contour1} for the benchmark values $M_{Z'}=1$ TeV and $g_L^{\mu\mu}=1.137$. The green region represents the $1\sigma$ constraint from $B_s-\bar B_s$ mixing, while the blue, orange, and purple contours correspond to the $1\sigma$, $2\sigma$, and $3\sigma$ regions, respectively, obtained from the $b\to s\mu^+\mu^-$ observables. An important feature of this analysis is that the latter observables predominantly favour a negative value of $\mathrm{Re}(\lambda_{bs})$, whereas the imaginary component is allowed on either side of zero. The intersection of these constraints identifies the region of the complex coupling plane compatible with both flavor sectors. The resulting best-fit point is located at $\lambda_{bs}=(-9.584\times10^{-4}, 2.345\times10^{-3})$. Thus, the combined flavor constraints provide a quantitative determination of the allowed complex $b-s$ coupling and establish the parameter region in which the $Z'$-mediated contributions can be consistently investigated. Similarly, for the scenario in which only $C_9$ receives new physics contributions, we constrain the complex $\lambda_{bs}$ parameter space using the bounds from $B_s-\bar B_s$ mixing and $b\to s\mu^+\mu^-$ observables. The overlap of the corresponding allowed regions identifies the parameter space consistent with both flavor sectors, thereby determining the preferred range of the complex $b-s$ coupling for the $Z'$-mediated new physics contributions in right panel of Fig. \ref{2D:contour1}.

On the other hand, to constrain the new physics parameter space associated with the $b\to d$ transition, we perform the analysis in the complex $\lambda_{bd}$ plane, incorporating the constraints from $B_d-\bar B_d$ mixing and $b\to d\mu^+\mu^-$ observables. The overlap of the corresponding allowed regions identifies the parameter space compatible with both mixing and semileptonic decay constraints, thereby determining the preferred range of the complex $b-d$ coupling for investigating the $Z'$-mediated new physics contributions. This is shown in Fig. \ref{2D:contour2}.

\begin{figure*}[t]
\centering
\includegraphics[width=0.55\linewidth]{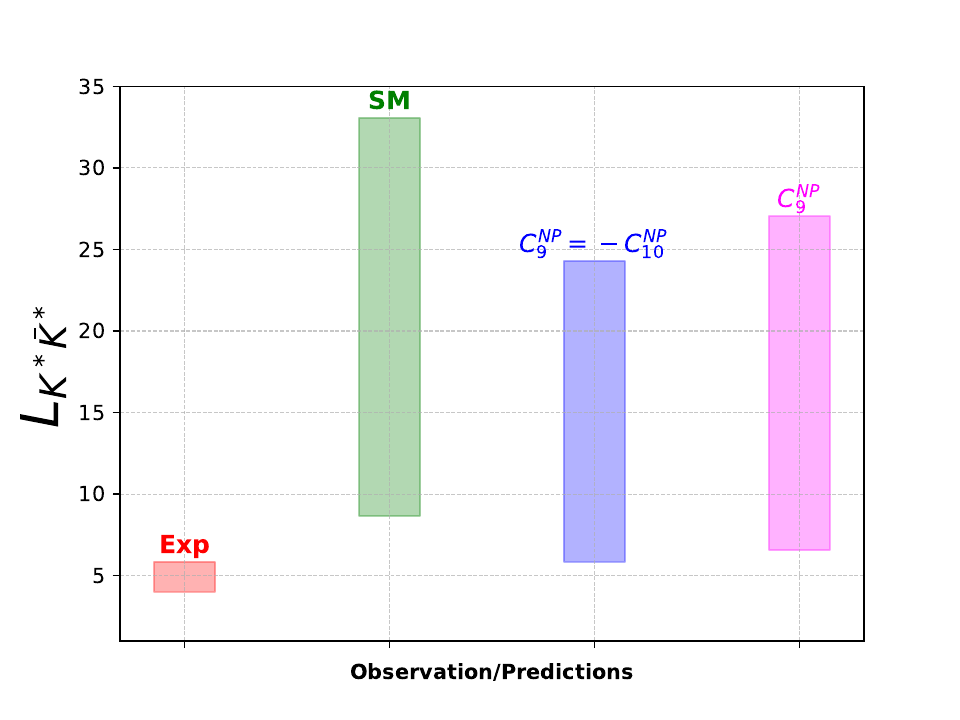}
\caption{Comparison of $L_{K^*\bar{K}^*}$ for the experimental measurement, the SM prediction, and the NP scenarios $C_9^{NP}=-C_{10}^{NP}$ and $C_9^{NP}$. The vertical bands represent the corresponding ranges of the observable.}
\label{Final_flavor}
\end{figure*}



\subsection{Impact of new physics on the $L_{K^{*} \bar{K}^{*}}$ puzzle}
The non-leptonic observable $L_{K^{*}\bar{K}^{*}}$, associated with the $B_{s(d)}\to K^{*}\bar{K}^{*}$ decays, provides an avenue to investigate the potential effects of physics beyond the Standard Model (SM) in the flavor sector. Motivated by the discrepancy between the experimental measurement and the SM prediction, we investigate the impact of NP contributions to the $b\to s(d)\mu^+\mu^-$ transitions on this observable. In particular, we consider two NP scenarios, namely $C_9^{NP}=-C_{10}^{NP}$ and $C_9^{NP}$, and examine their implications for $L_{K^{*}\bar{K}^{*}}$ by comparing the corresponding predictions with the experimental measurement and the SM expectation. This analysis allows us to assess whether the considered NP scenarios can alleviate the discrepancy between the experimental result and the SM prediction.
The relevant amplitudes, following Ref.~\cite{Beneke:2006hg, Bartsch:2008ps}, are given as follows
\begin{align}
\mathcal{A}(\bar{B}_d \to \bar{K}^{*0}K^{*0})
&= \sum_{p=u,c} A_{\bar{K}^{*}K^{*}}
\Big[
\alpha_4^p - \frac{1}{2}\alpha_{4,\mathrm{EW}}^p
+ \beta_3^p + \beta_4^p\nonumber\\
&- \frac{1}{2}\beta_{3,\mathrm{EW}}^p
- \frac{1}{2}\beta_{4,\mathrm{EW}}^p
\Big]
\nonumber\\
&\quad + \sum_{p=u,c} A_{K^{*}\bar{K}^{*}}
\Big[
\beta_4^p - \frac{1}{2}\beta_{4,\mathrm{EW}}^p
\Big],
\end{align}
\begin{align}
\mathcal{A}(\bar{B}_s \to \bar{K}^{*0}K^{*0})
&= \sum_{p=u,c} \lambda_p^{(s)}
\Bigg\{
A_{\bar{K}^{*}K^{*}}
\Big[
\beta_4^p - \frac{1}{2}\beta_{4,\mathrm{EW}}^p
\Big]
\nonumber\\
&\qquad\quad + A_{K^{*}\bar{K}^{*}}
\Big[
\alpha_4^p - \frac{1}{2}\alpha_{4,\mathrm{EW}}^p
+ \beta_3^p + \beta_4^p\nonumber\\
&- \frac{1}{2}\beta_{3,\mathrm{EW}}^p
- \frac{1}{2}\beta_{4,\mathrm{EW}}^p
\Big]
\Bigg\}.
\end{align}
\begin{figure*}[t]
    \centering

    \begin{minipage}[t]{0.45\textwidth}
        \centering
        \includegraphics[height=50mm,width=65mm]{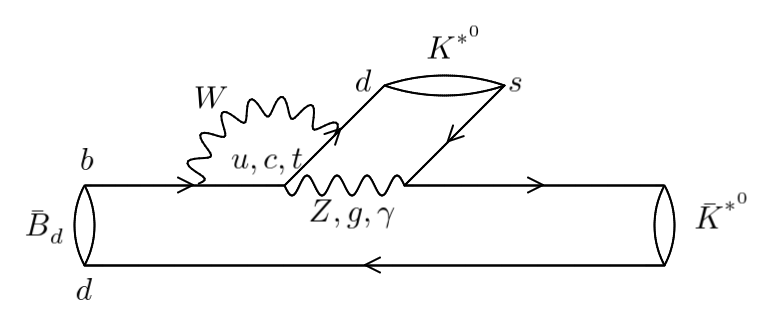}
    \end{minipage}
    \begin{minipage}[t]{0.45\textwidth}
        \centering
        \includegraphics[height=50mm,width=65mm]{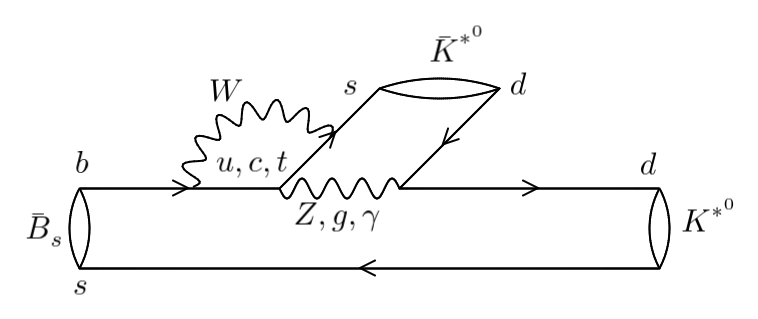}
    \end{minipage}

\caption{SM contributions to the non-leptonic decays $\bar{B}_{d,s}\to K^{*0} \bar{K}^{*0}$.}
   \label{fig:kstarkstar}  
\end{figure*}
\begin{table*}[ht]
\centering
\scriptsize
\renewcommand{\arraystretch}{1.8}
\setlength{\tabcolsep}{18pt}
\begin{tabular}{|c|c|c|c|c|}
\hline
Observable & Exp. \cite{LHCb:2025ftm} & SM & $C_9^{NP}=-C_{10}^{NP}$ & $C_9^{NP}$ \\ \hline
$L_{K^*\bar{K}^*}$ & $4.92 \pm 0.55 \pm 0.48 \pm 0.02 \pm 0.10$ & $20.87 \pm 12.21$ & $15.07 \pm 9.23$ & $16.80 \pm 10.25$ \\ \hline
\end{tabular}
\caption{$\mathcal{L}_{K^*\bar{K}^*}$ between experimental data, the SM prediction, and the NP scenarios.}
\label{tab:num}
\end{table*}
Here, the $\alpha$ parameters incorporate the contributions from vertex corrections, penguin diagrams, and hard-spectator scattering, whereas the $\beta$ parameters account for the weak-annihilation contributions. The normalization factors for the two configurations, $A_{\bar{K}^{*}K^{*}}$ and $A_{K^{*}\bar{K}^{*}}$, are identical and are given by
\begin{equation}
A_{\bar{K}^{*}K^{*}} = A_{K^{*}\bar{K}^{*}} =
\frac{G_F}{\sqrt{2}}\,m_{B_q}^{2}\,f_{K^{*}}\,
A^{B_q\to K^{*}}(0),
\end{equation}
where $G_F$ denotes the Fermi constant, $m_{B_q}$ is the mass of the 
parent $B_q$ meson, $f_{K^{*}}$ represents the decay constant of the 
$K^{*}$ meson, and $A^{B_q\to K^{*}}(0)$ is the corresponding 
transition form factor evaluated at zero momentum transfer.
In order to quantify the relative decay rates of the $\bar{B}_s$ and $\bar{B}_d$ 
mesons into the $K^{*0}\bar{K}^{*0}$ final state, the defined the observable 
$\mathcal{L}_{K^{*}\bar{K}^{*}}$ \cite{Biswas:2023pyw} is given as
\begin{align}
\mathcal{L}_{K^{*}\bar{K}^{*}}
&= \rho(m_{K^{*0}},m_{\bar{K}^{*0}})
\frac{\mathcal{B}(\bar{B}_s\to K^{*0}\bar{K}^{*0})}
{\mathcal{B}(\bar{B}_d\to K^{*0}\bar{K}^{*0})}
\frac{f_{B_s}^{L}}{f_{B_d}^{L}}
\end{align}
where $\rho(m_1,m_2)$ accounts for the ratio of the phase-space factor, and is expressed as
\begin{align}
\rho(m_1,m_2)
&= \frac{\tau_{B_d}}{\tau_{B_s}}
\frac{m_{B_s}^3}{m_{B_d}^3}
\nonumber\\
&\quad\times
\frac{
\sqrt{\big[m_{B_d}^2-(m_1+m_2)^2\big]
\big[m_{B_d}^2-(m_1-m_2)^2\big]}
}{
\sqrt{\big[m_{B_s}^2-(m_1+m_2)^2\big]
\big[m_{B_s}^2-(m_1-m_2)^2\big]}
}.
\end{align}
The relevant meson masses and lifetimes are taken 
from the PDG~\cite{ParticleDataGroup:2026mpi}.
From the definition of $\mathcal{L}_{K^{*}\bar{K}^{*}}$, we now evaluate this observable within the SM and in the presence of the two NP scenarios considered in our analysis, namely $C_9^{\mu \mu, \rm NP}=-C_{10}^{\mu \mu, \rm NP}$ and $C_9^{\mu \mu, \rm NP}$. The corresponding predictions are obtained using the allowed parameter ranges from the $b\to s(d)\mu^+\mu^-$ transitions, and are compared with the experimentally measured value to examine the extent to which these NP contributions can accommodate the observed discrepancy.

The comparison presented in Fig. \ref{Final_flavor} reveals a pronounced deviation between the experimentally observed value of $\mathcal{L}_{K^*K^*}$ and its SM prediction. In particular, the SM allowed range lies substantially above the experimental interval, indicating a considerable tension between the theoretical expectation and the measured result. We further examine the implications of two representative NP scenarios, characterized by $C_9^{\mu \mu, \rm NP}=-C_{10}^{\mu \mu, \rm NP}$ and $C_9^{\mu \mu, \rm NP}$, respectively. Interestingly, the inclusion of NP contributions leads to a downward shift in the predicted range relative to the SM expectation, thereby reducing the discrepancy with the experimental observation. Among the two scenarios, the $C_9^{\mu \mu, \rm NP}=-C_{10}^{\mu \mu, \rm NP}$ hypothesis yields a comparatively lower allowed range than the scenario involving $C_9^{\mu \mu, \rm NP}$ alone. Nevertheless, both predictions remain above the experimentally preferred interval, with no apparent overlap between the corresponding ranges. This observation suggests that, although the considered NP contributions improve the agreement with experimental data, these benchmark scenarios do not, by themselves, provide a complete resolution of the observed discrepancy. The result highlights the potential importance of exploring additional NP contributions or more general combinations of Wilson coefficients to achieve a better description of the experimental observable.
\section{Conclusion}
The $U(1)_{B-3L_\mu}$ gauge symmetry is anomalous in the SM spectrum and is made anomaly free by adding three right-handed neutrinos with charges $(-3,a,-a)$, taking $a=0$ for simplicity. The scalar $\Phi$ that breaks this symmetry gives mass to the $Z'$ and also generates the Majorana masses of the right-handed neutrinos.
In the flavour sector, we have investigated the implications of the $Z'$-mediated new physics contributions to the $b\to s\mu^+\mu^-$ and $b\to d\mu^+\mu^-$ transitions, taking into account the constraints from the corresponding neutral $B_s-\bar B_s$ and $B_d-\bar B_d$ mixing observables. By performing a numerical analysis in the complex $\lambda_{bs}$ and $\lambda_{bd}$ parameter planes, we have identified the allowed regions of the new physics couplings for different Wilson coefficient scenarios, namely, $C_9^{\mu \mu, \rm NP}$ alone and $C_9^{\mu \mu, \rm NP}=-C_{10}^{\mu \mu, \rm NP}$. These constraints provide an important framework for exploring the impact of the underlying new physics interactions on other flavour-changing neutral-current processes.

In particular, we have examined the implications for the $B\to K^*\bar K^*$ decay observables using the parameter space allowed by the $b\to s(d)\mu^+\mu^-$ transitions. This provides an opportunity to investigate the possible correlations between semileptonic and nonleptonic $B$-meson decays within the same new physics framework. The resulting predictions can be tested against future experimental measurements, offering an independent avenue to examine the consistency of the proposed $Z'$ scenario. Further improvements in the precision of the branching fractions and polarization observables, together with better control over hadronic uncertainties, will be particularly important in distinguishing the Standard Model expectations from possible new physics effects. Future measurements at Belle II and LHCb will therefore play an important role in further probing the flavour structure of the model and testing the correlations among different $B$-meson decay channels.
The type-I seesaw mechanism and the texture of the mass matrices follow directly from the same gauge $U(1)_{B-3L_\mu}$ symmetry. Using the CI parametrization, a scan over the complex orthogonal matrix $\mathbf{R}$ yields benchmark points consistent with NuFIT~6.1 data for a hierarchical RHN spectrum $(10^{11},10^{12},10^{13})$~GeV. The same $Z'$ that is tied to neutrino mass generation couples to quarks through $B$ and to muons through $-3L_\mu$, and it is this coupling that gives the $b\to s(d)\mu^+\mu^-$ transitions and their implications for $B\to K^*\bar K^*$. The $U(1)_{B-3L_\mu}$ gauge boson thus provides a common origin for neutrino masses and mixing and for the flavor observables studied here.
\section{Acknowledgement}
DP and PM would like to acknowledge the support of the Prime Minister's Research Fellowship, Government of India. 

\appendix
\section{Parameters in Neutrino sector}
\label{App:A}

\subsection{Unitary Matrices from Charged Lepton Sector}

\begin{equation}
\begin{aligned}
V^\ell_L &=
\begin{pmatrix}
0.839905 & 0 & -0.542733 \\
0 & -1 & 0 \\
0.542733 & 2.22045\times10^{-16} & 0.839905
\end{pmatrix},
\\[0.5cm]
V^\ell_R &=
\begin{pmatrix}
-0.939361 & 0 & -0.342931 \\
0 & -1 & 0 \\
0.342931 & 0 & -0.939361
\end{pmatrix}.
\end{aligned}
\end{equation}

\subsection{Unitary matrix for the diagonalization of the right-handed neutrino mass matrix}

\[
U_M =
\resizebox{\columnwidth}{!}{$
\begin{pmatrix}
0.18612 & 0.288029 & -0.939361 \\
-1.01159 - 0.118335 i & 0.363472 - 0.362275 i & -0.0889824 - 0.134528 i \\
0.315802 - 0.379055 i & 0.96681 + 0.136197 i & 0.359017 - 0.0333428 i
\end{pmatrix}
$}
\]

\subsection{Benchmark values of Yukawa couplings from neutrino sector}

\begin{table*}[ht]
\centering
\scriptsize
\renewcommand{\arraystretch}{1.3}
\begin{tabular}{|c|c|c|}
\hline
\textbf{BPs} & \textbf{Yukawa Matrix $Y$} & \textbf{$R$ Matrix} \\
\hline

BP-1
&
$\begin{pmatrix}
0.201+0.067i & 0.037+0.220i & 0.021-0.015i \\
0.001+0.019i & -0.008+0.011i & 0.005-0.006i \\
-0.035-0.070i & 0.065-0.069i & -0.006-0.008i
\end{pmatrix}$
&
$\begin{pmatrix}
-0.162-1.628i & -0.665-0.322i & -1.831+0.261i \\
2.554-1.425i & -0.880-1.957i & -1.308-1.465i \\
-1.723-1.959i & -2.132+0.909i & -0.887+1.624i
\end{pmatrix}$
\\
\hline

BP-2
&
$\begin{pmatrix}
-0.127+0.130i & 0.294+0.598i & 0.137-0.458i \\
-0.054+0.002i & -0.046+0.212i & 0.136-0.085i \\
-0.000-0.096i & -0.350-0.100i & 0.134+0.228i
\end{pmatrix}$
&
$\begin{pmatrix}
-0.596+0.693i & 1.525-0.540i & 0.929+1.332i \\
2.628+1.735i & 3.855-5.163i & 4.519+3.395i \\
-1.728+2.399i & 5.176+4.005i & -3.668+4.521i
\end{pmatrix}$
\\
\hline

BP-3
&
$\begin{pmatrix}
0.099-0.072i & 0.098-0.239i & -0.096+0.047i \\
0.033+0.004i & 0.060-0.019i & -0.019-0.015i \\
-0.014+0.060i & 0.041+0.138i & 0.037-0.054i
\end{pmatrix}$
&
$\begin{pmatrix}
0.120-0.588i & -1.609+0.737i & -0.974-1.289i \\
1.616+0.080i & -1.399-0.347i & 0.326-1.884i \\
-0.053+1.130i & -0.402-1.744i & 2.288-0.281i
\end{pmatrix}$
\\
\hline

\end{tabular}

\caption{Benchmark points showing the Yukawa matrices $Y$ and the corresponding complex orthogonal $R$ matrices.}
\label{tab:benchmark_Y_R}
\end{table*}


\bibliographystyle{utphys28mod}
\bibliography{MPDR}

\end{document}